\documentstyle[aps,pre,epsf,epsfig]{revtex}
\include{epsf}
\preprint{}
\begin{document} 
\title{The conditional tunneling time for reflection using the WKB wave-function}
\author{S. Anantha Ramakrishna\footnote{Presently at The Blackett Laboratory, Imperial College, London SW7 2BZ, U.K.}$^{(a)}$ and A.M. Jayannavar$^{(b)}$\footnote{E-mail: jayan@iopb.res.in}}
\address{$^{(a)}$Raman Research Institute, C.V. Raman Avenue, Bangalore 560 080, India.\\
$^{(b)}$ Institute of Physics, Sachivalaya Marg, Bhubaneswar 751 005, India}
\maketitle
\date{\today}
\begin{abstract}
We derive an expression for the conditional time for the reflection of a wave from 
an arbitrary potential barrier using the WKB wavefunction in the barrier region.  
Our result indicates that 
the conditional times for transmission and reflection are equal for a symmetric
barrier within the validity of the WKB approach. \\

\end{abstract}

In this paper, we deal with the dynamical aspect of scattering, namely, the
time delay undergone by a
wave in the process of being scattered by a potential. Obviously, this time is
related to the actual time of sojourn in the scattering region. For the case of
a classical particle, both the times will be one and the same. But it is not
obvious that the equality would hold for a wave. This time of sojourn during  
the scattering is of interest for mesoscopic systems, 
 and has  a long and controversial 
history (See Ref. \cite{chiao,landauer94,landauer91,HandS,buttiker90} for 
recent reviews). The deformable nature of a wave-packet makes it difficult to 
accept the Wigner phase ($\phi$) delay time [$\tau_{w} = \hbar (\partial \phi)/
(\partial E)$]\cite{wigner}, based on following  the motion of a fiducial 
feature such as 
the peak on the wave-packet, as causally related to the actual time of sojourn
in the spatial region of interest\cite{buttiker82,landauer92}. Hence, several 
authors have made other proposals for identifying a physically {\it meaningful}
timescale of interaction of the quantum particle with the scattering potential. 
These include the quantum clocks that
utilize the co-evolution, in a locally applied infinitesimal field / potential,
of an extra degree of freedom (such as the spin \cite{buttiker83}) attached to
the traversing particle. Even 
these proposals are not completely free from problems 
\cite{landauer94,HandS,golub90,sarsojourn}.

One strongly debated issue, among others, has been the conditional scattering 
time, conditional upon the incoming and outgoing channels. Thus, we have the 
conditional transmission or reflection time in the one-dimensional case. The
case of transmission (tunneling for sub-barrier energies) has been dealt with
extensively by several researchers using different methods\cite{buttiker82,buttiker83,bruinsma,jayan87,leavens,SandB,buttiker85,martin93}. There
is near unanimous agreement  that the traversal time for tunneling across a 
nearly opaque barrier of width $L$ and height $V_{0}$ is given by the 
B\"{u}ttiker-Landauer time\cite{buttiker82}:
\begin{equation}
\tau_{BL} = \frac{mL}{\hbar \kappa} ,
\end{equation}
where m is the mass of the particle and $\hbar\kappa = \sqrt{2m(V_{0}-E)}$. This can,
of course, be generalized to an arbitrary potential as an integral over 
infinitesimally wide rectangular barriers, if the law of addition of 
scattering times for non-intersecting spatial regions holds. 
But there is no such universal 
agreement about the time for reflection. In this paper, we will use the WKB 
wave-function in the barrier region for a tunneling particle to obtain an 
expression for the conditional time of reflection from an arbitrary potential. 
In an earlier work\cite{jayan87}, this approach
was successfully used  to determine the traversal time in agreement with the
B\"{u}ttiker-Landauer time. Our current 
work indicates that the conditional time for 
Reflection is the same as the conditional traversal time for a symmetric 
barrier within the limits of the WKB approximation.\\      

Let us consider an arbitrary (static) potential $V(x)$ as shown in Fig.~1, 
with the wave at an energy $E$ incident from the left. The wave is now 
partially transmitted to the right with a probability amplitude $T$, and 
partially reflected to the left with a probability amplitude $R$. Within the 
barrier region the wave function $\psi(x)$ of the particle is a superposition
of growing and decaying real functions (which are purely exponential for a
rectangular barrier). The current density at any point, $j(x) = 
(\hbar /2im)[ \psi^{*} (d\psi /dx) - \psi (d\psi^{*}/dx)]$, is non-zero only
due to the complex coefficients differing by a phase, in the superposition of
real functions. Now one can associate a total velocity field $v(x)$ with the 
particles, given by $j(x) = v(x)P(x)$, where $P(x)=\psi^{*}(x) \psi(x)$ is the
probability density. This turns out to be the same the velocity field obtained
in the Bohmian view\cite{bohm,spiller}.
Using the WKB wave-function in the barrier region\cite{landau},
\begin{equation}
\psi(x)=\frac{A}{\sqrt{p(x)}} \left[ \frac{1}{2}\exp\left( -\int_{x}^{a}
\frac{p(x')}{\hbar}dx' \right) - i \exp \left( \int_{x}^{a} \frac{p(x')}{\hbar}
dx' \right) \right]  ,
\end{equation}
where $p(x)=\sqrt{2m(V(x)-E)}$ and $A$ is a complex normalization coefficient, 
one can evaluate the velocity field $v(x)$:
\begin{eqnarray}
v(x) &=& \frac{p(x)}{m} \left[ \frac{1}{4} \exp\left( -2\int_{x}^{a}\frac{p(x')}
{\hbar}dx' \right) + \exp \left( 2 \int_{x}^{a} \frac{p(x')}{\hbar} dx' \right)
\right]^{-1}, \\
 & \simeq & \frac{p(x)}{m} \exp\left( -2\int_{x}^{a}\frac{p(x')} {\hbar}dx' 
\right).
\end{eqnarray}
The latter approximation is valid because $\int p(x)/\hbar dx \gg 1$ inside the 
barrier, which is the essence of the WKB approximation. 

\begin{figure}
\epsfxsize=230pt
\begin{center}{\mbox{\epsffile{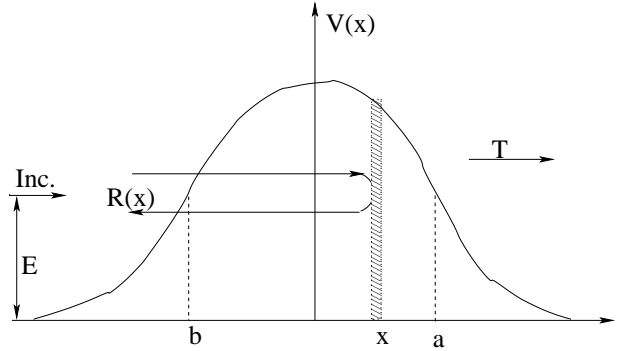}}}
\end{center}
\caption{A schematic of the potential showing the classical turning points and
the partial reflections in the WKB approach.
\label{WKBpot}}
\end{figure}

Now at any point, $x$, inside the barrier, the forward (moving to the right) 
particles would be reflected backwards (to the left) due to the potential 
barrier extending from $x$ to $a$ with a probability ${\cal R}(x)$ (in this WKB
picture there are no multiple scatterings). By noting that the speed at a 
given point would only depend on the potential at that point,  
the total velocity field can be decomposed into a forward and a 
backward velocity field:
\begin{equation}
v(x) = \frac{p(x)}{m} - {\cal R}(x)\frac{p(x)}{m}={\cal T}(x)\frac{p(x)}{m}.
\label{totalvelocity_eq}
\end{equation}
Now in Eq.~(4), the exponential term is nothing but the transmission probability (${\cal T}(x)$)
for the barrier extending from $x$ to $a$ in the WKB approximation.  Hence the 
forward velocity was identified as $p(x)/m$, giving the traversal time as
\begin{equation}
\tau^{(T)}= \int_{b}^{a} \frac{m}{p(x')} dx'  .
\end{equation}
For a rectangular barrier, this reduces to the B\"{u}ttiker-Landauer time.
It should be noted that such an identification of the forward velocity field
holds within the WKB picture and might not possible in general. We also point
out that the form of $v(x)$ in Eq. \ref{totalvelocity_eq} naturally leads to 
a simple interpretation in the form of a symmetric statistical composition 
law for the velocity field without any quantum interference terms. 

Now we will  give  an expression for the conditional reflection time within the
WKB picture. The reflection time for tunneling can be defined as a properly
weighted sum  over the transit time of the particles partially reflected from 
each point within the barrier as  
\begin{equation}
\tau^{(R)} = \int_{b}^{a} 2 \left[ \int_{b}^{x}
\frac{m~dx'}{p(x')} \right] \frac{{ \cal R}(x)~dx}{\int_{b}^{a} {\cal R} (x)
dx}. \end{equation}
In other words, the particle reflected at a point $x$ (see Fig.~1) will take a 
time of $2\int_{b}^{x}p(x')/m~dx'$ to go upto the point $x$ and back. 
${\cal R}(x)$ is the (probability) reflection coefficient of the barrier
extending from only $x$ upto $a$, and in the WKB approximation is given by
\cite{bohm} \begin{equation}
{\cal R}(x) = 1 - \exp \left[ -2 \int_{x}^{a} p(x'')/\hbar ~dx'' \right].
\end{equation}
Now Eqs.~(7) and (8) constitute the complete expression for the conditional 
reflection time.

Next, we will proceed to calculate the reflection time for 
two symmetric potentials (there exists a centre of reflection), 
{\it viz.}, a rectangular potential barrier
of height $V_{r}$ and width $L$, and a parabolic potential barrier
$V(x) = -(1/2)~ \omega^{2} x^{2}$. For the rectangular barrier, we obtain the
reflection delay time as
\begin{equation}
\tau^{(R)} = 
\frac{mL}{p_{0} N} - \frac{m \hbar}{p_{0}^{2} N} + \frac{2m}{p_{0}N}
(\frac{\hbar}{2p_{0}})^{2}
\left(1 - e^{-2p_{0}L/\hbar} \right),
\end{equation}
where $N = [1+\hbar/2p_{0}L (1-e^{-2p_{0}L/\hbar})]$ and $p_{0} =
\sqrt{2m(V_{r}-E)}$. We note that for a sufficiently wide barrier $p_{0}L/\hbar
\gg 1$, and the reflection time can be expanded in powers of $\hbar/p_{0}$.  To
the zeroeth order, $ \tau^{(R)} = \tau^{(T)} = mL/p_{0}$, {\it i.e.}, the
reflection time is the same as the transmission time.  In the case of the
parabolic barrier, the transmission time is $\tau^{(T)} = \pi \sqrt{m}/ \omega$
and the reflection time is 
\begin{equation}
\tau^{(R)} = \frac{\pi \sqrt{m}}{\omega} + \frac{2 \sqrt{m}}{\omega}
\int_{-2E/\sqrt{\omega}}^{2E/\sqrt{\omega}} \frac{ \sin^{-1} \left(\frac{\omega
x'}
{2E} \right) {\cal R}(x')
~dx'}{\int_{-2E/\sqrt{\omega}}^{2E/\sqrt{\omega}}{\cal R}(x') ~dx'}
\end{equation} where
\begin{equation}
{\cal R}(x) = 1 - \exp \left( \frac{-\pi E \sqrt{m}}{\hbar \omega} \right) \exp
\left[ \frac{2E\sqrt{m}}{\hbar \omega} \left( \frac{\omega x}{\sqrt{2E} }
\sqrt{1-\omega^{2} x^{2}/2E} + \sin^{-1} (\omega x/\sqrt{2 E}) \right) \right]
\end{equation}
${\cal R}(x) \sim 1$ for a reasonably small $\omega$ or a broad potential. Then
the second part of the expression~(10) for $\tau^{(R)}$ is negligible, giving
$\tau^{(R)} = \tau^{(T)}$. For ${\cal R}(x) <1$, the reflection time is slightly
lesser than the transmission time. Thus within the validity of the WKB approach,
the reflection and the transmission times are roughly 
equal in this case also.

We now discuss the significance of our result for the reflection time for a 
symmetric barrier by comparing with other known timescales. The Wigner phase 
delay time is the same for reflection or traversal for a symmetric barrier, and
does not distinguish between reflected and transmitted particles. The 
conditional reflection and traversal times calculated by using a local 
infinitesimal imaginary potential as a clock\cite{buttiker90,golub90,sarsojourn,sardelay} can also be shown to be equal for any arbitrary  symmetric 
potential. The spin rotation time ($\tau_z$) 
defined by B\"{u}ttiker\cite{buttiker83}, 
 however, yields  different reflection
and transmission times in general for any potential (symmetric or otherwise), 
although the spin precession times $\tau_{y}$ for 
reflection and transmission are equal.  
But it should be noted that there is a
relation between the z-components of the spin of the transmitted and the reflected waves 
due to
conservation of angular momentum. This makes it difficult to define separate
conditional reflection and transmission times using the  spin-rotation
($\tau_{z}$). 

In conclusion, we have given an expression for the conditional reflection 
time for barrier tunneling using the WKB wave-function in the barrier region. 
The conditional reflection time appears to be the same as the conditional 
traversal time for a symmetric barrier within the validity of the WKB picture.

The authors would like to thank Prof. N. Kumar for stimulating discussions.
SAR would like to acknowledge the Institute of Physics, Bhubaneswar for 
hospitality during a visit when this work was carried out.

\end{document}